\newcommand{\bea}{\begin{eqnarray}}
\newcommand{\eea}{\end{eqnarray}}
\newcommand{\ba}{\begin{array}}
\newcommand{\ea}{\end{array}}
\newcommand{\be}{\begin{equation}}
\newcommand{\ee}{\end{equation}}
\newcommand{\bc}{\begin{center}}
\newcommand{\ec}{\end{center}}

\newcommand{\Eqref}[1]{Eq.(\ref{#1})}
\newcommand{\eqref}[1]{eq.(\ref{eqn:\LBL:#1})}

\newcommand{\Eqsref}[2]{Eqs.(\ref{#1},\,\ref{#2})}
\newcommand{\eqs}[2]{\mbox{Eqs.(\ref{#1},\ref{#2})}}
\newcommand{\eq}[1]{\mbox{Eq.(\ref{#1})}}
\newcommand{\refc}[1]{Ref.\cite{#1}}

\newcommand{\cL}{{\cal L}}
\newcommand{\cO}{{\cal O}}

\newcommand{\non}{\nonumber}
\newcommand{\inv}[1]{\frac{1}{#1}}
\def\lsim{\; \raise0.3ex\hbox{$<$\kern-0.75em
      \raise-1.1ex\hbox{$\sim$}}\; }
\def\gsim{\; \raise0.3ex\hbox{$>$\kern-0.75em
      \raise-1.1ex\hbox{$\sim$}}\; }

\newcommand{\bfT}{{\bf\rm\bfT}}
\def\id{\leavevmode\hbox{\small1\kern-3.3pt\normalsize1}}

\newcommand{\del}{\partial}
\newcommand{\ra}{\rangle}
\newcommand{\la}{\langle}
\newcommand{\rar}{\rightarrow}
\newcommand{\apj}[3]{Ap.~J.~{{\bf #1} {(#2)} {#3}}}
\newcommand{\apjl}[3]{Ap.~J.~Lett.~{{\bf #1} {(#2)} {#3}}}

\newcommand{\pr}[3]{Phys.~Rev.~{{\bf #1} {(#2)} {#3}}}

\newcommand{\prl}[3]{Phys.~Rev.~Lett.~{{\bf #1} {(#2)} {#3}}}

\newcommand{\prep}[3]{Phys.~Rep.~{{\bf #1} {(#2)} {#3}}}

\newcommand{\jmp}[3]{J.~Math.~Phys.~{{\bf #1} {(#2)} {#3}}}

\newcommand{\np}[3]{Nucl.~Phys.~{{\bf #1} {(#2)} {#3}}}

\newcommand{\lbmeff}{\cL_{\rm eff}^{\beta,\mu}}
\newcommand{\lvaceff}{\cL_{\rm eff}^{\rm vac}}

\newcommand{\lcal}{\mbox{${\cal L}$}}

\newcommand{\G}{\mbox{${\Gamma}$}}
\newcommand{\bb}{\bar{B}}
\newcommand{\bm}{\bar{M}}

\newcommand{\bbeta}{\bar{\beta}}
\newcommand{\bmu}{\bar{\mu}}
\newcommand{\brho}{\bar{\rho}}
\newcommand{\baf}{\bar{f}}
\newcommand{\ga}{\gamma}
\newcommand{\gl}{\begin{array}{c}{\scriptstyle <}
\\[-2.7ex]{\scriptstyle >}
                 \end{array}}
\newcommand{\AUTHORS}{\
{\centering
{\large Per Elmfors,\footnote{Email address:
elmfors@nordita.dk.}}\raisebox{1ex}{,a}
{\large Per
Liljenberg,\footnote{Email address: tfepl@fy.chalmers.se.}}\raisebox{1ex}{,b}
{\large David Persson\footnote{Email address:
tfedp@fy.chalmers.se.}}\raisebox{1ex}{,b}\\
 and {\large Bo-Sture Skagerstam\footnote{Email address:
tfebss@fy.chalmers.se. Research supported by the Swedish National
Research Council under contract no. 8244-311.}}\raisebox{1ex}{,b,c}
 \\[5mm]
{\sl \raisebox{1ex}{a}NORDITA,
   Blegdamsvej 17,
 DK-2100 Copenhagen \O, Denmark \\ }
{\sl \raisebox{1ex}{b}Institute of Theoretical Physics,
   Chalmers University of Technology and \\
    University of G\"oteborg, S-412 96 G\"oteborg,
Sweden \\ }
{\sl \raisebox{1ex}{c}University of Kalmar,
Box 905, S-391 29 Kalmar, Sweden\\[10mm]}
}}
\documentstyle [12pt]{article}
\begin{document}
\thispagestyle{empty}
\begin{flushright}
            G\"{o}teborg ITP 94--12 \\
June 1994\\
 \end{flushright}
\begin{center}
\baselineskip 1.2cm
{\Huge\bf  Condensation and Magnetization \\
of the Relativistic  Bose Gas}\\[10mm]
\normalsize
\end{center}
%
\AUTHORS
%
\begin{abstract}
\normalsize
\noindent
We  present a simple proof
of the absence of  Bose--Einstein condensation of a relativistic boson gas,
in any finite local magnetic field in less than five dimensions.
We show that the relativistic charged boson gas exhibit a genuine
 Meissner--Ochsenfeld effect
of the Schafroth form at fixed supercritical density.
As in the well--known non--relativistic case, this total expulsion of a
magnetic field is caused by the condensation of the Bose gas at vanishing
magnetic field.  The result is discussed in the context of kaon condensation in
  neutron stars.
 \end{abstract}
\newpage
\setcounter{footnote}{0}
\setcounter{page}{1}
\noindent
In this letter we study some aspects of  an ideal
 charged Bose gas at finite temperature $T$ and chemical potential $\mu$  in
  presence of a static uniform magnetic field.
Magnetic fields $B$  associated with  compact astrophysical
objects may range between $B={\cal O}
(10^{4})\;$T for magnetic white dwarfs  to $B={\cal O}
(10^{10})\;$T for supernovae \cite{ginzburg91etc}. As a reference we
recall that the  characteristic magnetic field in QED is
$m_e^2/e={\cal O}(10^9)\;$T.
For such systems one may expect that also  thermal effects are of
importance.
In the absence of a magnetic field it is known that the relativistic charged
 boson gas exhibits a Bose--Einstein condensation
(see e.g. Ref.\cite{haber&weldon81}). In the presence of a magnetic field the
non--relativistic charged boson gas was studied by Schafroth~\cite{Schafroth}.
The relativistic system has recently been extensively
 considered by Daicic et  al.~\cite{DaicicFGK93}.
However, we do not agree with their conclusions about the relativistic
Meissner--Ochsenfeld effect.
\bigskip\\
According to \refc{Fradkin}, introducing an external field is
equivalent to introducing an external current independent of the dynamics of
the system considered. Including the term $\cL_{\rm ext}=
	j^\nu_{\rm ext} A_\nu$ in the classical Lagrangian for scalar QED,
Euler--Lagrange's equation of motion for $A_\nu$
reads
\be
  \del^\mu F_{\mu\nu}= j_\nu + j_\nu^{\rm ext}~~~,
	\label{eq-eulera}
\ee
where $j_\nu$ is the induced current.
When quantizing the particles but not  the gauge field we may, in the uniform
magnetostatic case, perform the
functional integral over the scalar field, and write the effective Lagrangian
 density as
\be
  \cL_{\rm eff}= \cL_0 +\lvaceff  +\lbmeff +
		\cL_{\rm ext}~~~,
\ee
where $\cL_0=  -\frac12 B^2$ is the tree level term,
$\lvaceff$ is the one--loop vacuum correction, and $\lbmeff$ is the
thermal contribution. In terms of the (average microscopic)
magnetic induction ${\bf B}=\nabla \times {\bf A}$, and the external magnetic
 field $ {\bf H}$, such that  $\nabla \times  {\bf H}={\bf j}_{\rm ext}$,
we may (upon neglecting a surface term) write $\cL_{\rm ext}={\bf B}
\cdot {\bf H}$. The mean-field equation then follows from a minimization of
the effective action
 (in natural units $\alpha=e^2/4\pi$)
\be
 	{\bf B}= {\bf H} +{ \bf M}({\bf B})~~~,
\label{eq-bhfield}
\ee
where the average microscopic  magnetization is defined by
\be
  M_i(x)= \frac\del{\del B_i(x)} (\lvaceff +\lbmeff)~~~,
	\label{eq-magdef}
\ee
such that the expectation value of the induced current is
 $\la {\bf j} \ra= \nabla \times  {\bf M}$. Considering the external field
as the acting field (i.e. the field felt by the particles)
as in \refc{DaicicFGK93}, means that the induced current
is neglected in \eq{eq-eulera}, and  lead to erroneous
conclusions about the relativistic Meissner effect. In the case of Meissner
effect, the
magnetization appears to be stronger than the external field, as explained
below, and thus certainly may not be neglected. In the
original work by Schafroth~\cite{Schafroth} and also in \refc{May},
 $B$ was used as the acting field,
consistent with \eqs{eq-bhfield}{eq-magdef}.
 The contribution from the
vacuum polarization $\lvaceff$ to the magnetization is negligible for small
magnetic fields, and of no importance when considering the topics
discussed here, so it will be neglected in what follows.
\bigskip\\
In \refc{elm&per&ska93a} the thermal part of the effective Lagrangian
 $\lcal_{\rm eff}^{\beta,\mu}$ for
the relativistic charged boson gas in a homogeneous magnetic field, neglecting
all boundary effects,  was
 calculated and found to be related to the free energy $F$,
and grand partition
function  $Z(B,T,\mu)$, according to
$\lcal_{\rm eff}^{\beta,\mu} =  \log Z/V \beta = - F/V$.
Generalizing to  $d+1$ dimensional spacetime~\cite{May}, the free
energy density $f_d=F_d/V_d$ is
\be
	f_d =  -\frac{eB}{2^{d-1}\pi^{d/2} \G(\frac{d}{2})}
	\sum_{n=0}^{\infty} \int_{0}^{\infty} dp
	\frac{p^{d-1}}{E_n(p)}\biggl(
	f_B^{+}(E_n(p))+f_B^{-}(E_n(p))\biggr)~~~,
\label{eq-fd}
\ee
where $f_B^{\pm}$ are the one-particle distributions
$f_B^{\pm}(\omega ) = (e^{\beta(\omega \mp \mu)}-1)^{-1}$,
and the energy is given by
\be
   E_n(p) =\sqrt{m^2+ p^2+(2n+1)eB}~~~.
\ee
Following the steps in Ref.\cite{elm&per&ska93a}, separating the
particle($+$) and antiparticle($-$) contributions, and introducing
dimensionless
quantities $ (\bbeta \equiv m\beta,\; \bmu \equiv \mu/m,\;
 \baf_d \equiv f_d/m^{d+1},\;
	 \brho_d \equiv \rho_d/m^d,\; \bb \equiv eB/m^2,\;
 \bar{H} \equiv eH/m^2)$,
we rewrite \eq{eq-fd} as
\be
\baf_{d}^{\pm} =- (\frac{1}{4\pi})^\frac{d+1}{2}
 \sum_{k=1}^{\infty} \int_{0}^{\infty} \frac{ds}{s}
 s^{-(\frac{d+1}{2})} \exp(- s - \frac{\bbeta^2 k^2}{4s})
 \frac{\bb s}{\sinh \bb s}
e^{\pm k \bbeta \bmu}~~~.
\ee
The integral and sum here are absolutely convergent, but if we expand in
 powers of $\bb$, the series is only asymptotic, and as $\bmu$ assumes its
critical value at the lowest energy level $\bmu_c=E_0(0)/m=\sqrt{1+\bb}$, the
coefficients   become divergent.
However, we may still in a simple manner discuss the analytical behaviour of
the free energy and the magnetization,  using the following inequality
\be\label{ineq}
 (1+2\bb s)e^{-\bb s}>\frac{\bb s}{\sinh \bb s} > (1+\bb s)e^{-\bb s}~~~,
\ee
which we write as
\be
\frac{\bb s}{\sinh \bb s}\gl(1+c_0 \bb s)e^{- \bb s} ,~~~c_0\in [1,2]~~~,
\ee
and similarly for the  purpose of the
magnetization
 \be
	\frac{\partial}{\partial \bb} \frac{\bb s}{\sinh \bb s} \gl
	-  c_1 \bb s^2 e^{-\bb s},~~~~~c_1\in [\frac{1}{3},2]~~~.
\ee
Introducing  the function ($x=\bbeta k s$)
\be
	\ga^{\pm}(x) \equiv \bbeta[\frac{\bmu_c^2}{x}  (x-x_0)^2 +
		(\bmu_c \mp \bmu) ]~~~,~ x_0\equiv 1/(2\bmu_c)~~~,
\ee
and identifying a  Jonqui\`ere's function with an exponential argument
$
\psi_{a}(z) \equiv \sum_{k=1}^{\infty}  e^{-k z}/k^a
$,
the  charge density $ e\brho = e( \brho^+ + \brho^-)=
		-e \frac{\partial \baf}{\partial \bmu}$, and the
magnetization
$\bm_d =- 4\pi\alpha  \frac{\partial \baf_d }{\partial \bb}$ are written
\bea
	\brho_{d}^{\pm}&\gl&
	\pm(\frac{1}{4 \pi \bbeta})^\frac{d+1}{2}\bbeta
	\int_{0}^{\infty} \frac{dx}{x} x^{-(\frac{d+1}{2})}
	\left\{ \psi_{\frac{d-1}{2}}[\ga^\pm(x)] +
	 c_0 \bb  \bbeta x \psi_{\frac{d-3}{2}}[\ga^\pm(x)]
 	\right\}~~~,
\label{eq-brho} \\
	\bm_{d}^{\pm}&\gl&-4\pi \alpha\: c_1 \bb
		 (\frac{1}{4 \pi \bbeta})^\frac{d+1}{2}
	 \bbeta^2\int_{0}^{\infty} \frac{dx}{x} x^{-(\frac{d+1}{2})}
	x^2 \psi_{\frac{d-3}{2}}[\ga^\pm(x)]~~~.
\label{eq-magn}
\eea
The charge density naturally splits into two parts
$\brho \equiv \brho_{\rm reg} +\brho_{\rm div}$, where $\brho_{\rm reg}$,
the first term in \eq{eq-brho}, for a
vanishing magnetic field is the charge density of noncondensed states,
and $\brho_{{\rm div}}$ is  the  second  term in \eq{eq-brho}.
We shall now investigate $\brho_{\rm div}$ for $\bmu \rar \bmu_c$, in order to
see if the magnetized Bose gas can form a condensate.
The leading behaviour of $\psi_{\frac{d-3}{2}}(\ga^+)$ close to $x=x_0$ is
 for $\frac{d-3}{2} \leq 1$, or $d$ even~\cite{Schafroth}
\be
	\psi_{\frac{d-3}{2}}[\ga^{\pm}(x)] \sim \Gamma(\frac{5-d}2 )
	\left\{ \bbeta [ \frac{\bmu_c^2}{x}(x-x_0)^2 +
	(\bmu_c \mp \bmu)]\right\}^{\frac{d-5}{2}}~~~.
\ee
For positive chemical potential ($\bmu<0$ can be treated similarly)
we thus see that
$\brho_{d,{\rm div}}^{-}$   remains finite, whereas at $\bmu=\bmu_c$,
$\psi_{\frac{d-3}{2}}[\ga^{+}(x)] \sim (x-x_0)^{d-5}$ so that
$\brho_{d,{\rm div}}^{+}$  diverges for $d\leq 4$. Actually, also
$\brho_{d,{\rm reg}}^{+}$  diverges for $d\leq 2$. At
finite magnetic field, for an arbitrary density
$\brho_{d\leq 4}$ there is thus a value of the chemical potential $\bmu$ for
 any inverse
temperature $\beta$, such that $\brho_d = \brho_{d,{\rm reg}} +
\brho_{d,{\rm div}}$, hence the
magnetized Bose gas does not condense for $d\leq 4$ (and in the absence of the
field the
 gas does not condense for $d \leq 2$). Actually this can be seen in a
physically
 more illuminating way. The divergent contribution to the charge density should
come
from the lowest Landau level ($n=0$). Separating out that level we have
according to \eq{eq-fd}, after a change of variables of summation and
integration
\bea
	\brho_{d}^{+}& =& -\frac{\bb}{2^{d-2}\pi^{d/2}\Gamma(\frac{d}{2}-1)}
	\int_{0}^{\infty}dx x^{d-3} \frac{1}
	{\exp[\bbeta(\sqrt{1+x^2+\bb}-\bmu)]-1}\nonumber \\
	&& -\frac{\bb}{2^{d-2}\pi^{d/2}\Gamma(\frac{d}{2}-1)}
	\sum_{n=0}^{\infty}\int_{0}^{\infty}dx x^{d-3} \frac{1}
	{\exp[\bbeta(\sqrt{1+x^2+(2n+3)\bb}-\bmu)]-1}~.
\eea
The first term is easily seen to diverge at $\bmu=\bmu_c$ for $d\leq 4$, and
the second term is exactly of the same form as if the lowest Landau level was
included,
but with the lowest energy ($\sqrt{1+3\bb}$) always larger than the chemical
potential ($\bmu\leq \bmu_c$), for finite $B$.
 It thus follows from the above inequalities
that this sum  always is finite.
 Hence even though no true
Bose--Einstein condensate can form for $d\leq 4$,
 the lowest Landau level can play the role of the
groundstate and accommodate a large charge density.
\bigskip\\
Let us now consider the physically most  relevant case of $d=3$,
suppress the dimensional index and return to dimensionful quantities.
In the case of vanishing magnetic field $\rho_{{\rm div}}\equiv 0$, and
$ \rho_{{\rm reg}}$ is finite as $\mu \rar \mu_c(B=0)=m$.
At fixed temperature there is thus a critical density
\be
	\rho_c(T)=\rho_{\rm reg}(\mu=m,T,B=0)=
	\frac{Tm^2}{\pi^2} \sum_{k=1}^{\infty}
	\frac{1}{k}K_2(k\beta m) \sinh(k\beta m)~~~.
\label{eq-rhocrit}
\ee
Condensation   occurs if this critical density is exceeded.
In the limit $\mu \rightarrow \mu_c$ the integrals in
\Eqsref{eq-brho}{eq-magn} are dominated by $x$ close to $x_0$,
and we obtain the leading behaviour ($\mu_c=\sqrt{m^2+eB}$)
\bea
\label{rholead}
\rho_{\rm div}^+ &\gl& \frac{1}{4\pi\sqrt{2}}  c_0eB\,T
 \sqrt{\frac{\mu_c}{\mu_c-\mu}}~~~, \\
\label{Mlead}
	M &\gl& -
	\frac{e}{8\pi\sqrt{2}} c_1 eB \frac{T}{\mu_c}
	\sqrt{\frac{\mu_c}{\mu_c-\mu}}~~~.
\eea
We now whish to consider the magnetization in the limit of vanishing magnetic
field $B \rar 0$, at fixed supercritical density $\rho>\rho_c$, i.e
when a condensate is formed. Then we must have that $\bmu \rar \bmu_c$, and
in this limit $\brho_{\rm reg} \rar \brho_c$, so that \Eqsref{rholead}{Mlead}
give that the magnetization is approaching a constant. The obtained
magnetization law
\be
	M(B\rar 0) = -\frac{e}{2m}\frac{c_1}{c_0}(\rho-\rho_c)  ~~~,
\label{eq-bnollmag}
\ee
  is   exactly of the Schafroth form~\cite{Schafroth},
who derived it in the non--relativistic case with the constant $c_1/c_0=1$.
We may actually determine the limiting values of $c_0$ and $c_1$ here,
 using the previous method of
separating out the lowest Landau level.
Taking the limit $B\rar 0$ for all higher Landau levels we find the
supercritical density
\bea
	\rho&=&\frac{eB}{2\pi^2}\int_0^\infty dp
	\inv{\exp[\beta(E_0(p)-\mu)]-1} \non\\
	&&+\inv{2\pi^2}\int_0^\infty dp\,p^2
	\left.\Bigl(f^+_B(\sqrt{p^2+m^2})+
	f^-_B(\sqrt{p^2+m^2})\Bigr)\right|_{\mu=m}~~~,
\eea
and the magnetization
\bea
	M&=&-\frac{e^2B}{4\pi^2}\int_0^\infty\frac{dp}{E_0(p)}
	\inv{\exp[\beta(E_0(p)-\mu)]-1}\ .
\eea
In the $n=0$ term we let $\mu$ approach
$\sqrt{m^2+eB}$ in such a way that a prescribed
$\rho$ is obtained. We thus find  $c_0=c_1=2$,
in agreement
with Shafroth's~\cite{Schafroth} non--relativistic result.
In  the ultra--relativistic $(T\gg m)$ and non--relativistic
$(T\ll m)$ limit, \eq{eq-rhocrit} gives the well--known result
\be
	\rho_c\approx  \left\{
	\ba{ll}
	\frac{1}{3}T^2m&,~T\gg m \\
	\zeta(\frac{3}{2})(\frac{1}{2\pi})^\frac{3}{2}
	(Tm)^{3/2}&,~T\ll m
\ea \right.~~~.
\ee
Below the critical temperature
we can therefore write the magnetization
\be
	M(B\rar 0) \approx -\frac{e}{2m}\rho\left\{
	\ba{ll}{}
 	[1-(T/T_c)^2]&  ,~T\gg m \\
 	{}[1-(T/T_c)^{3/2}]& ,~T\ll m
	\ea \right. ~~~.
\label{M}
\ee
where $T_c$ is defined by the condition that $\rho_c(T=T_c)=\rho$.
There is thus   a critical field $H_c\equiv -M(B\rar 0)$, such
that for external fields smaller than $H_c$ \eq{eq-bhfield} has no
solution, and the field is expelled from the Bose gas. This is the
well--known Meissner--Ochsenfeld effect. The expulsion of the
field is caused by surface currents, and since we do not consider surface
effects here, they   appear as external currents in our formalism.
The origin of the Meissner--Ochsenfeld effect is here, as well as in the
non-relativistic case, that the total free energy is minimized if a condensate
 is formed, which is  only   possible in vanishing magnetic field,
and is not connected to the high temperature pair production as claimed in
\refc{DaicicFGK93}.
 The perfect
expulsion of the externally applied  field increases the free energy per unit
volume by~\cite{Schrieffer} $H_{\rm appl}^2/2$ , which causes
penetration at supercritical field strengths.
If we instead  keep $\mu=m$ fixed as $B \rar 0$,
\eq{Mlead} gives  ($\mu_c \simeq m+eB/2m$)
\be
\label{eq-sqrtmagn}
	M \simeq  -e\frac{c_1}{8\pi}\sqrt{eB}\,T\ .
\ee
The square root magnetization
law in Eq.(\ref{eq-sqrtmagn}) was obtained in \refc{DaicicFGK93}\footnote{Here
an expansion of the form $\mu=m+g(\beta,\beta_c,B)$, for $g\ll eB/m$ was
used. In the limit $B \rar 0$ this is only valid for $g \equiv 0$, and
thus this corresponds to keeping the chemical potential (and not the density as
was intended) fixed for small
magnetic fields.}, where the corresponding  $c_1=6\sqrt2\pi \,\zeta[-1/2,1/2]
\simeq 1.6 $ was found.
Considering the externally applied field as the acting field made the authors
of \refc{DaicicFGK93} draw erroneous conclusions about the Meissner effect
from \eq{eq-sqrtmagn}. We claim that the magnetization law of
\eq{eq-bnollmag}, leading to genuine Meissner--Ochsenfeld effect,
 may be derived from the expressions in \refc{DaicicFGK93}
if a supercritical density is correctly held fixed.
\bigskip\\
There has recently been some discussion about condensation of $K^-$ mesons in
the core of neutron stars and its influence on the equation of state
\cite{BrownBTPL94}. Since there are also very large magnetic fields in some
neutron stars it is interesting to ask whether the field can influence the
condensation.
If magnetic flux is trapped in the inner core when the protons become
superconducting it is believed that flux tubes are formed with field strengths
much  higher than on the surface. The details of the formation and dynamics of
such flux tubes are not known so we shall only estimate some relevant
quantities related to the kaons.
We can here safely put $T=0$ compared to the kaon effective mass $m^*_K\simeq
210\;$MeV and use \Eqref{M} to compute the critical field strength. For the
typical value $\rho_K\simeq 0.1\;$fm$^{-3}$ we find $H_c\simeq 10^{12}\;$T
which is far above the maximal observed fields on the surface of neutron stars,
but comparable with the field strength of the flux tubes in the proton
condensate.  The kaons are essentially non--relativistic so we can use
Schafroth's \cite{Schafroth} formula for the penetration depth $d=(m_K/4\pi
e\rho_K)^{1/2}\simeq 3\;$fm. Let us estimate the order of magnitude of the
field gradient in the flux tube by $H_c/d$. Such a field gradient exerts an
enormous force even on small magnetic dipoles. For instance, we estimate the
force on the electron and the neutron, due to their intrinsic magnetic moments
$\mu_e$ and $\mu_n$, to be $\cO(10^3)\;$N and $\cO(1)\;$N, respectively.

\vspace{3mm}
\begin{center}
{\bf ACKNOWLEDGEMENT}
\end{center}
\vspace{3mm}
One of the authors (B.-S.~S.) thanks
 NFR for providing the financial support. We are
grateful to P. Mazur, F. Wilczek and V. Thorsson for providing
us with information on
dense nuclear matter in neutron stars.
 \vspace{3mm}
%
%

%
\end{document}